\documentclass[aps,prl,preprint,showpacs]{revtex4-2}
\usepackage{graphicx}
\usepackage{helvet}
\usepackage{amsmath,amssymb}
\usepackage{bm}

\begin{document}

\title{Reduced superfluid density in a doped spin liquid candidate}

\author{K. Wakamatsu$^{1}$}
\author{Y. Ueno$^{1}$ }
\author{K. Miyagawa$^{1}$ }
\author{H. Taniguchi$^{2}$ }
\author{K. Kanoda$^{1,\ast}$ }

\affiliation{
$^{1}$Department of Applied Physics, University of Tokyo, Bunkyo-ku, Tokyo, 113-8656, Japan\\}
\affiliation{
$^{2}$Graduate School of Science and Engineering, Saitama University, Saitama 338-8570, Japan\\}

\date{\today}

\begin{abstract}
	A quantum spin liquid (QSL) would be an exotic stage for superconductivity. 
	A promising candidate for a doped QSL is the organic triangular-lattice system, 
$\kappa$-(BEDT-TTF)$_{4}$Hg$_{2.89}$Br$_{8}$, 
which hosts a non-Fermi liquid and magnetism of a QSL nature and shows superconductivity 
upon cooling. 
	Here, we report that its superfluid density is anomalously reduced, indicating 
the existence of substantial incoherent spectral weight and weak superconducting phase stiffness.
	Moreover, the ratio of the superconducting transition temperature to the nominal Fermi 
temperature is as large as 0.1, orders of magnitude beyond typical BCS values.
	These observations in a system free from competing orders that complicate the similar 
issue in underdoped cuprates give a clue to the enigmatic missing superfluid density in 
doped Mott insulators.
\\
\end{abstract}


\maketitle

%{INTRODUCTION}
%\section{Introduction}

	The superfluid density, a spectral weight contributing to superconducting condensates, 
determines the fundamental nature of superconductivity, such as phase stiffness \cite{Emery_1995}, 
and thus closely pertains to phase fluctuations. 
	A quantum spin liquid (QSL) is a nontrivial magnetic state free from order and possibly bearing 
fractionalization, entanglement and/or chirality \cite{Zhou_2017, Savary_2017, Broholm_2020}. 
	The superconductivity that emerges by doping such an anomalous magnetic state 
\cite{Anderson_1987, Lee_2006, Powell_2011, Jiang_2020, Jiang_2021} is of keen interest with 
regard to the unconventional pairing nature out of a BCS framework. 
	In the absence of adequate model materials that address this question, 
the layered organic conductor with a triangular lattice, 
$\kappa$-(BEDT-TTF)$_{4}$Hg$_{2.89}$Br$_{8}$ ($\kappa$-HgBr) \cite{Oike_2017}, 
has recently arisen as a promising doped QSL candidate. 
	The present work experimentally investigates the nature of superconductivity that emerges 
in this system in terms of the superfluid density, 
which is a key parameter to elucidate the possible unconventionality of pairing.

\begin{figure}
	\includegraphics[width=12cm]{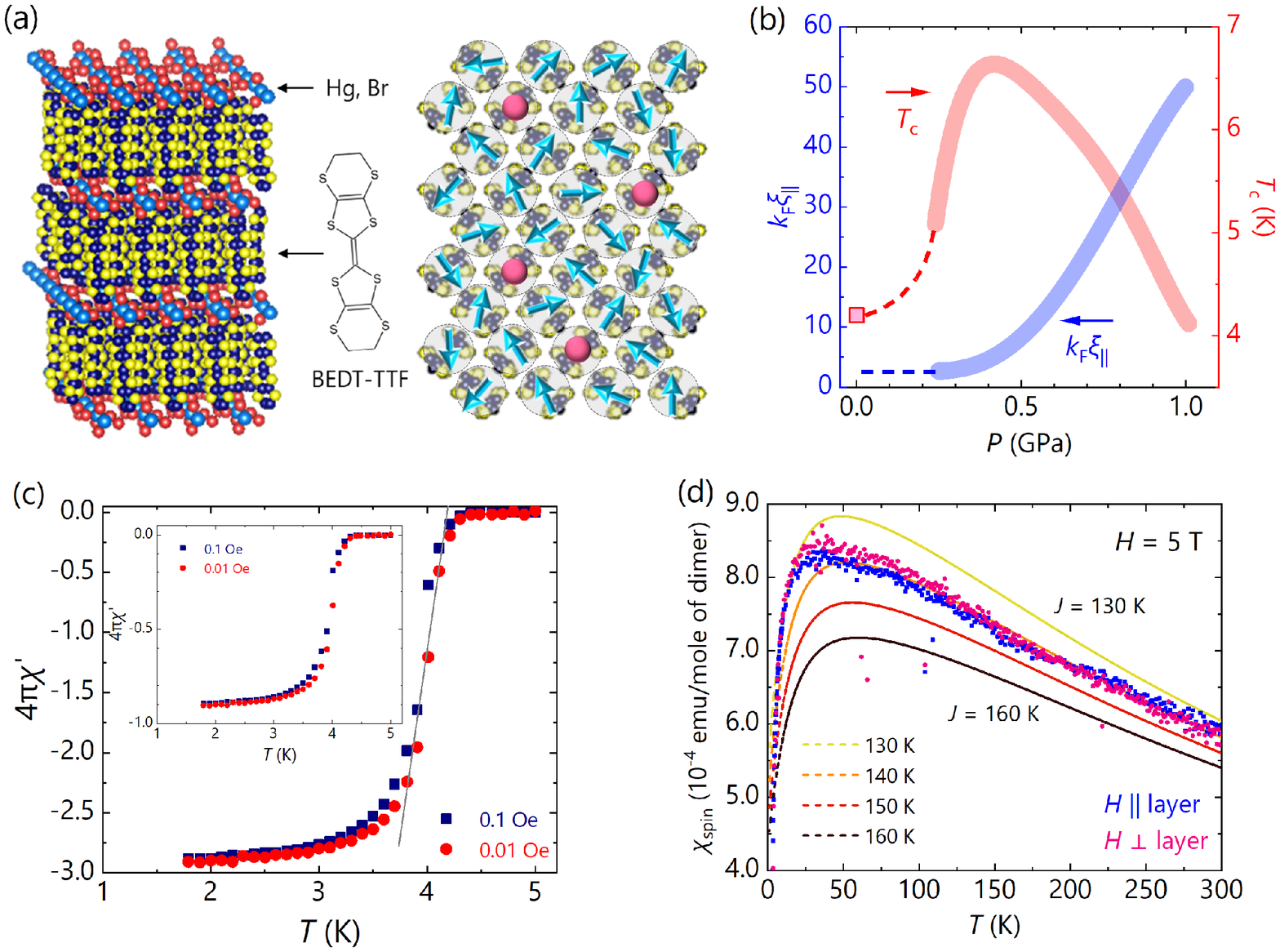}
	\caption{\label{Fig1} (Color online) Crystal structure and superconducting properties of 
$\kappa$-HgBr. 
(a) Crystal structure of $\kappa$-HgBr. 
BEDT-TTF molecules and anions (Br and Hg) constitute conducting and insulating layers, respectively. 
In the conducting plane, two BEDT-TTF molecules form a dimer, and 
the dimers form a triangular lattice. In the case of $\kappa$-HgBr, band filling deviates from 
half-filling, and 11\% hole doping is realized. 
	The blue arrows and red spheres depict singly occupied holes with spins and 
vacancies (holons) produced by the doped holes. 
	(b) Pressure dependence of $T_{\textrm{c}}$ and $k_{\textrm{F}}\xi_{\parallel}$. 
	$k_{\textrm{F}}$ and $\xi_{\parallel}$ express the Fermi wavenumber and the in-plane superconducting coherence length, respectively. The data under pressure are taken from Ref. \cite{Suzuki_2022}, 
and $T_{\textrm{c}}$ at ambient pressure is for the sample used in this study. 
	$T_{\textrm{c}}$ forms a dorm against pressure, and the value of 
$k_{\textrm{F}}\xi_{\parallel}$ at low pressures is as low as $\sim$3, 
which means bare overlapping of the Cooper pairs. 
	(c) ac susceptibility of the $\kappa$-HgBr crystal used in the present study. 
	The ac fields with amplitudes of 0.01 and 0.1 Oe and a frequency of 10 Hz were applied 
perpendicular to the conducting layers. 
	The inset shows the ac susceptibility corrected for the demagnetization effect by 
approximating the crystal shape with a spheroid, which yields the demagnetization coefficient, 
$N$$\sim$0.69. 
	The magnitude of the corrected ac susceptibility inevitably contains an ambiguity of 
10-20\% considering the forcible approximation of the sample shape. 
	Thus, low-temperature saturation can be regarded to indicate complete diamagnetism. 
	(d) Temperature dependence of the spin susceptibility of $\kappa$-HgBr. 
	A magnetic field $H$ = 5 T was applied parallel and perpendicular to the conducting layers. 
	The core diamagnetism is corrected. This behavior reproduces the results reported in 
Ref. \cite{Oike_2017}.
}
\end{figure}

	$\kappa$-HgBr has a quasi-two-dimensional structure with insulating and conducting layers 
alternately stacked, and both layers have monoclinic sublattices 
\cite{Li_1998, Lyubovskaya_1987}. 
	In the conducting layers, the ET molecules form dimers, which constitute a quasi-triangular 
lattice characterized by two kinds of transfer integrals, $t$ and $t'$, between the adjacent 
dimer molecular orbitals (Fig. \ref{Fig1}(a)). 
	According to the molecular orbital calculations, the ratio, $t'/t$, in $\kappa$-HgBr is 
1.02 \cite{Shimizu_2011}, which means a nearly right triangular lattice. 
	Although most $\kappa$-(ET)$_{2}X$ compounds have half-filled bands, or equivalently, 
one hole carrier per dimer because of the monovalent anion $X^{-1}$, the band filling in 
$\kappa$-HgBr deviates from half by 0.055 assuming Hg$^{+2}$ and Br$^{-1}$ because of 
the nonstoichiometric composition of Hg, 2.89; if it is 3.0, the band filling will be half. 
	The compositional ratio of $4.0:2.89:8.0$ that results from the incommensurate sublattice of 
Hg against ET and Br is precisely determined by x-ray diffraction \cite{Li_1998, Lyubovskaya_1987} 
and cannot be varied. 
	The band-filling deviation from half corresponds to 11\% hole doping to the half-filled band and is consistent with Raman spectroscopy \cite{Yamamoto_2005}.

	Several experiments indicate that $\kappa$-HgBr is a hole-doped version of the QSL candidate, 
$\kappa$-(BEDT-TTF)$_{2}$Cu$_{2}$(CN)$_{3}$, hosting a similar triangular lattice and 
comparable or larger electron correlation \cite{Shimizu_2011}. 
	The resistivity and Hall coefficient \cite{Oike_2015} show that $\kappa$-HgBr hosts a 
non-Fermi liquid with the effective carrier density largely reduced from that of the total band 
electrons, suggesting that strong electron correlation prohibits double occupancy 
as if mobile carriers are restricted to doped holes. 
	Interestingly, the spin susceptibility of $\kappa$-HgBr is nearly perfectly scaled to that of 
$\kappa$-(BEDT-TTF)$_{2}$Cu$_{2}$(CN)$_{3}$. 
	$\kappa$-HgBr is thus regarded as a hole-doped version of 
$\kappa$-(BEDT-TTF)$_{2}$Cu$_{2}$(CN)$_{3}$. 
	Notably, superconductivity that appears at low temperatures is suggested to be of 
an unconventional nature out of the BCS picture in that the Cooper pairs are barely overlapping, 
saliently robust to the magnetic field and preformed prior to their condensate 
\cite{Suzuki_2022, Imajo_2021}, reminiscent of BEC-like pairing (Fig. \ref{Fig1}(b)). 
	The superfluid density ns of such unconventional condensates is of profound interest. 
	The present work investigates $n_{\textrm{s}}$ in this doped spin liquid candidate.

%\section{Results}
	We evaluated the $n_{\textrm{s}}$ value by measuring the in-plane magnetic penetration depth, 
$\lambda_{\parallel}$, whose inverse square is related to $n_{\textrm{s}}$ as 
$1/\lambda_{\parallel}^{2} = \mu_{0}n_{\textrm{s}}e^{2}/m^{\ast}$ ($\mu_{0}$: 
magnetic permeability, $n_{\textrm{s}}$: superfluid density, $e$: elementary charge, 
$m^{\ast}$: effective mass). 
	Among several experimental ways to estimate the $\lambda_{\parallel}$ value, we employed a method to evaluate it from the magnetization curve, as decribed later.
	The magnetization was measured with Quantum Design SQUID magnetometers (MPMS-XL and MPMS-5S), 
where a crystal was fixed to a plastic-straw sample holder with a Kapton film. 
	The net sample magnetization was obtained as the difference between the magnetizations with 
and without the sample. 
	The sample used in the present work is a rhombic single crystal with diagonal lengths of 
$\sim$2 and $\sim$2.3 mm, a thickness of $\sim$0.5 mm and a weight of $\sim$3 mg. 
	To date, the difficulty in obtaining crystals showing bulk superconductivity at 
ambient pressure has prohibited experimental access at ambient pressure \cite{Oike_2015}. 
	The present work at ambient pressure was, nevertheless, made possible by confirming 
the bulk superconductivity in the used crystal in terms of the ac susceptibility 
(Fig. \ref{Fig1}(c)). 
	Spin susceptibility in the normal state was also confirmed to 
reproduce the earlier results (Fig. \ref{Fig1}(d)).

\begin{figure}
\centering
	\includegraphics[width=15cm]{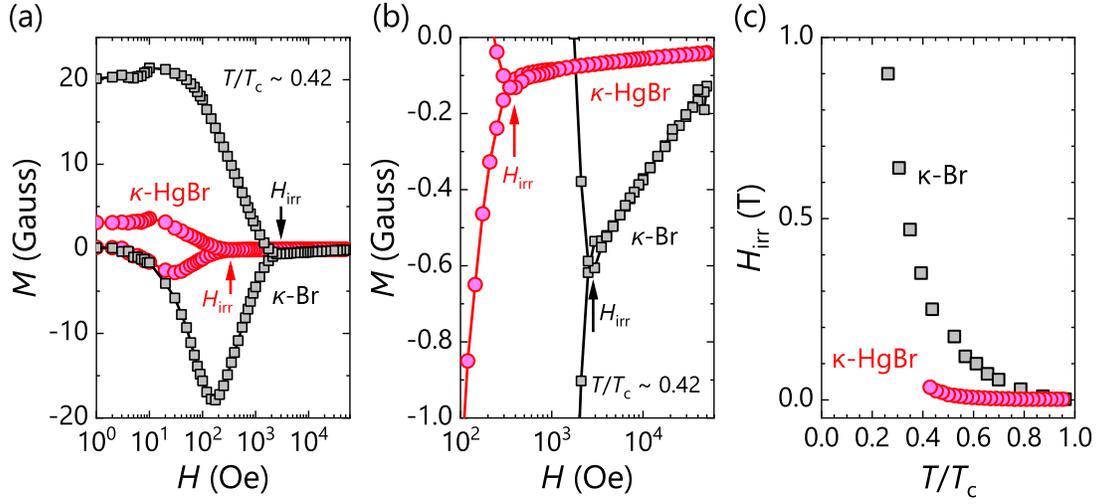}
	\caption{\label{Fig2} (Color online) Magnetization curves and irreversibility fields of 
$\kappa$-HgBr and $\kappa$-Br. 
	(a) Magnetization curves of $\kappa$-HgBr and $\kappa$-Br at the same reduced temperature, 
$T/T_{\textrm{c}}\sim$0.42. 
	The vertical axis is normalized by the sample volume. 
(b) Expanded panel of (a). 
(c) Irreversibility field versus $T/T_{\textrm{c}}\sim$ for $\kappa$-HgBr and $\kappa$-Br.
}
\end{figure}

\begin{figure}
\centering
	\includegraphics[width=15cm]{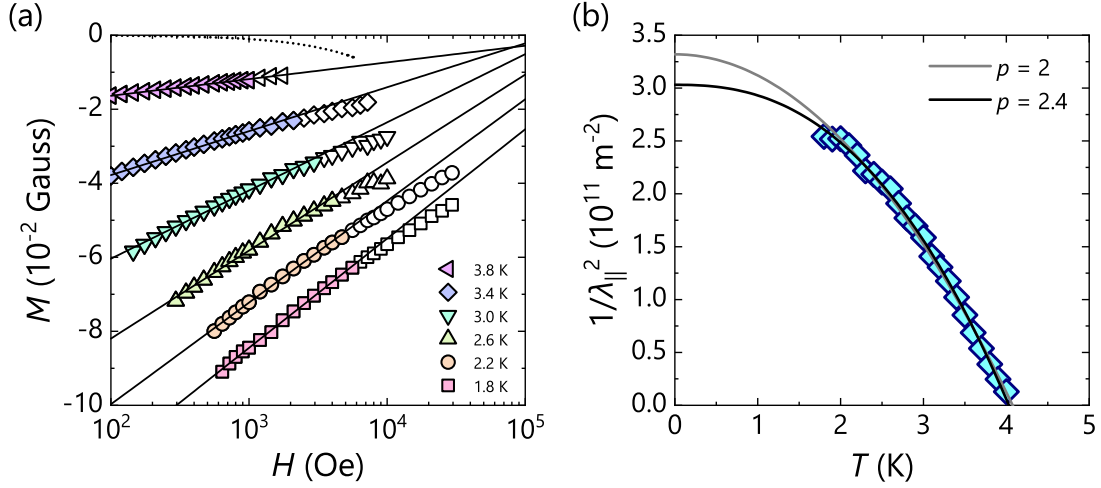}
	\caption{\label{Fig3} (Color online) Magnetization and penetration depth of $\kappa$-HgBr. 
(a) Magnetic field dependence of the magnetization in $\kappa$-HgBr in 
the reversible region at several temperatures. 
	The black lines are the fits of the formula 
$-4\pi M = (\alpha \phi_{0} / 8\pi \lambda _{\parallel}^{2}) \ln (\beta H_{\textrm{c2}\perp}/H)$ 
to the data. 
	The broken curve shows the calculated value of core diamagnetism. 
	To avoid the contribution of the core diamagnetism to the magnetization, 
we used the magnetization data (closed symbols) in the field range where the core diamagnetism is 
negligible in the analysis. 
	(c) Temperature dependence of the 1/{$\lambda_{\parallel}^{2}$} of $\kappa$-HgBr. 
	The line represents the fits of the form 
$1/\lambda_{\parallel}(T)^{2} = 
1/\lambda_{\parallel}(0)^{2} \times [1- (T/T_{\textrm{c}})^{p}]$ with 
$p$ as a free parameter, which yields 2.4 and is fixed at 2.
}
\end{figure}

\begin{figure}
	\includegraphics[width=15cm]{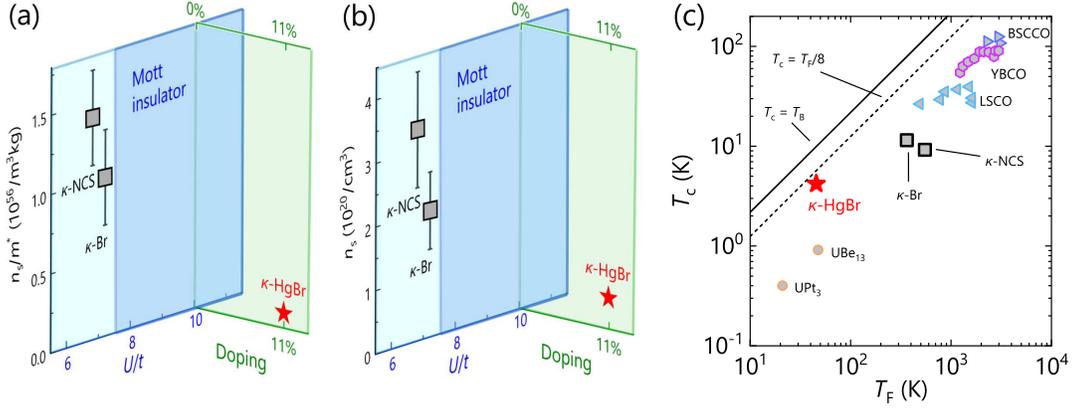}
	\caption{\label{Fig4} (Color online) Comparison of $n_{\textrm{s}}/m^{\ast}$ and 
$n_{\textrm{s}}$ between half-filled and hole-doped cases and Uemura plot in 
some strongly correlated materials.
	(a) and (b), Superfluid weight $n_{\textrm{s}}/m^{\ast}$ and superfluid density 
$n_{\textrm{s}}$ of $\kappa$-Br \cite{Wakamatsu_2020, Lang_1992, Yoneyama_2004, Andraka_1991, 
Taylor_2007}, $\kappa$-NCS \cite{Yoneyama_2004, Lang2_1992, Taylor_2007, Andraka_1989} and 
$\kappa$-HgBr. 
	The values of electron correlation $U/t$ refer to Ref. \cite{Shimizu_2011}. 
	(c) Superconducting transition temperature versus nominal Fermi temperature 
(so-called Uemura plot) for several strongly correlated electron systems of organic, 
cuprate and heavy electron materials. 
	The symbols of the red star and gray squares for $\kappa$-HgBr, $\kappa$-Br and 
$\kappa$-NCS are located using the present values, whereas other symbols for the cuprates and 
heavy electron systems are located using the information in 
Refs. \cite{Uemura_1989, Uemura_1991, Melo_1993, Hazra_2019}. BSCCO, YBCO and LSCO indicate 
the cuprates of 
Bi$_{\textrm{2-x}}$Pb$_{\textrm{x}}$Sr$_{\textrm{2}}$Ca$_{\textrm{2}}$Cu$_{\textrm{3}}$O$_{\textrm{10}}$, 
YBa$_{\textrm{2}}$Cu$_{\textrm{3}}$O$_{\textrm{y}}$, and 
La$_{\textrm{2-x}}$Sr$_{\textrm{x}}$CuO$_{4}$, respectively. 
	The line of $T_{\textrm{c}} =T_{\textrm{F}}/8$ indicates the two-dimensional limit of 
the BEC-like condensate.
}
\end{figure}

	$\lambda_{\parallel}$ can be deduced from the field dependence of the magnetization, 
$M(H)$, in the reversible field region under a magnetic field applied perpendicular to the layers. 
	The $M(H)$ for strong type II superconductors with 
$\kappa = \lambda_{\parallel}/\xi_{\parallel}\gg1$ ($\xi_{\parallel}$; 
the in-plane superconducting coherence length) is expressed by

\begin{equation}
-4 \pi M = (\alpha\phi_{0}/8\pi\lambda_{\parallel}^{2})\ln(\beta{H_{\textrm{c2}\perp}}/H)
\end{equation}

for a field range of 
$H_{\textrm{c1}\perp} \ll H_{\textrm{irr}} < H \ll H_{\textrm{c2}\perp}$ 
($H_{\textrm{irr}}$: irreversible magnetic field above which the magnetization is reversible, 
$H_{\textrm{c2}\perp}$: perpendicular upper critical field, 
$\phi_{0}$: flux quantum, $\alpha$: core contribution factor, 
$\beta$: a constant of the order of unity) \cite{DeGennes_1966}. 
	In this field range, vortices are free from collective pinning, and magnetization is 
reversible against field sweeps. 
	According to Eq. (1), $dM/d(\ln H)$ gives an estimate of $\lambda_{\parallel}$. 
	As the correction factor $\alpha$ from the London model, which does not consider 
the vortex core contribution, we employed $\alpha$ = 0.77, a value obtained in Ref. \cite{Hao_1991}.
	The $M(H)$ curves at fixed temperatures were measured under upward and downward field sweeps in 
the range of 0-5 T. 
	Fig. \ref{Fig2}(a) and \ref{Fig2}(b) shows the $M(H)$ curve of $\kappa$-HgBr at 
1.8 K along with that of the half-filled system $\kappa$-Br at 5 K 
(the same reduced temperature, $T/T_{\textrm{c}}\sim$0.42) for comparison. 
	$\kappa$-HgBr exhibits a peak in $M(H)$ at $H_{\textrm{p}}$ = 30 Oe and 
reversible behavior above $H_{\textrm{irr}}$ = 300 Oe; 
both the $H_{\textrm{p}}$ and $H_{\textrm{irr}}$ values are approximately an order of 
magnitude below those in $\kappa$-Br \cite{Wakamatsu_2020}, 
suggesting that the superconductivity in $\kappa$-HgBr is highly resilient to magnetic fields. 
	Accordingly, the temperature dependence of $H_{\textrm{irr}}$ shows that 
the vortex liquid phase is largely extended in the $H-T$ plane in $\kappa$-HgBr in comparison with 
$\kappa$-Br (Fig. \ref{Fig2}(c)). 
	The extended vortex liquid state in superconductivity with BEC-like character is 
theoretically suggested \cite{Adachi_2019}. 
	Magnetization curves above $H_{\textrm{irr}}$ at several temperatures are shown in 
Fig. \ref{Fig3}(a). 
	The gray lines indicate the fits of Eq. (1), 
and their slopes correspond to $(\alpha\phi_{0}/8\pi)[1/\lambda_{\parallel}(T)^{2}]$. 
	The fit also yields the values of $\beta H_{\textrm{c2}\perp}$, 
which is approximately 50 T at 2 K. 
	Conventionally, the $\beta$ value ranges from 1-2, e.g., 1.44 in the case of 
$\alpha$ = 0.77 \cite{Hao_1991}. 
	$H_{\textrm{c2}_{\perp}}$ of $\kappa$-HgBr at low temperatures is reported to be 
$\sim$15 T by torque measurements \cite{Imajo_2021} and largely exceeds 9 T 
(the highest field studied) by the Nernst effect, 
which is susceptible to vortex liquidity \cite{Suzuki_2022}, 
suggesting an extraordinarily large $H_{\textrm{c2}\perp}$ for $T_\textrm{c}$ = 4.2 K. 
	Nevertheless, the $\beta$ value in $\kappa$-HgBr likely exceeds the conventional range; 
probably, $\beta$ = 3-4. Although there is an argument on possible modifications of 
Eq. (1) \cite{Kogan_2006}, considering that the $\beta$ in $\kappa$-Br takes 
a conventional value approximately in the range of 1-2 \cite{Wakamatsu_2020, Kwok_1990}, 
the large $\beta$ value may characterize the peculiar nature of 
superconductivity in $\kappa$-HgBr.
	The temperature dependence of 1/$\lambda_{\parallel}(T)^{2}$ is shown in Fig. \ref{Fig3}(b). 
	The 1/$\lambda_{\parallel}(T)^{2}$ data are fitted by the formula 
$1/\lambda_{\parallel}(T)^{2} = 
1/\lambda_{\parallel}(0)^{2} \times (1-(T/T_{\textrm{c}})^{p})$ with exponent $p$, 
penetration depth $\lambda_{\parallel}(0)$ and $T_{\textrm{c}}$ as fitting parameters. 
	The gray line is the best fit with $p$ = 2.4, $\lambda_{\parallel}(0)$ = 1820 nm and 
$T_{\textrm{c}}$= 4.0 K and, for reference, the dotted line indicates a fit with 
$p$ fixed at the empirically adopted value, 2, 
which yields $\lambda_{\parallel}(0)$ = 1740 nm and $T_{\textrm{c}}$ = 4.0 K. 
	The $\lambda_{\parallel}(0)$ value in $\kappa$-HgBr is three to four times larger than 
those in the half-filled systems: 490-650 nm in $\kappa$-Br 
\cite{Wakamatsu_2020, Lang_1992, Yoneyama_2004} and 430-535 nm in $\kappa$-NCS 
\cite{Yoneyama_2004, Lang2_1992}. 
	With these $\lambda_{\parallel}(0)$) values, 
the relation $1/\lambda_{\parallel}^{2} = \mu_{0}n_{\textrm{s}}e^{2}/m^{\ast}$ yields 
$n_{\textrm{s}} / m^{\ast} = 9.4 \times 10^{54}$, $1.0(\pm 0.3)\times 10^{56}$, and 
$1.4(\pm0.3)\times10^{56}$ (m$^{3}$kg)$^{-1}$ in 
$\kappa$-HgBr, $\kappa$-Br and $\kappa$-NCS, respectively (Fig. \ref{Fig4}(a)). 
	Further assuming that $m^{\ast}$ is proportional to the specific heat $\gamma$, 
which is 55-57 mJ/K$^{2}$mol for $\kappa$-HgBr \cite{Imajo_2021, Naito_2005}, 22-28 mJ/K$^{2}$mol 
for $\kappa$-Br \cite{Andraka_1991, Taylor_2007} and 
25-35 mJ/K$^{2}$mol for $\kappa$-NCS \cite{Taylor_2007, Andraka_1989}, 
and considering the conducting plane cross-sectional area of the unit cell, 
one obtains $n_{\textrm{s}} = 4.9 \times 10 ^{19}$, $2.0 (\pm 0.6)\times 10^{20}$, 
and $3.3 (\pm 0.9) \times 10^{20}$ (cm)$^{-3}$ in $\kappa$-HgBr, $\kappa$-Br and 
$\kappa$-NCS, respectively (Fig. \ref{Fig4}(b)). 
	$\kappa$-Br is located very close to the Mott transition \cite{Kanoda_2006}, 
which may be responsible for the smaller $n_{\textrm{s}}$ value in 
$\kappa$-Br than in $\kappa$-NCS even in the half-filled case. 
	These values give the relative ratio of the superfluid density as 
$n_{\textrm{s}}(\kappa$-HgBr)/$n_{\textrm{s}}(\kappa$-NCS) = 15$\pm$4\% or 
$n_{\textrm{s}}(\kappa$-HgBr)/$n_{\textrm{s}}(\kappa$-Br) = 25$\pm$8\%. 
	Clearly, the superfluid density in $\kappa$-HgBr is significantly reduced from 
that of the half-filled systems, suggesting that only a fractional spectral weight approximately 
equivalent to the doped hole density contributes to the superconducting condensate. 
	This behavior is called gossamer superconductivity, 
which is argued to emerge in doped Mott insulators.

%
%
%
%
%
%
%

%\section{Discussion}
	The superfluid density is related to the phase stiffness, which is expressed as 
$\hbar^{2}d$/4$\mu_{0}e^{2}\lambda_{\parallel}^{2}$ 
($d$: interlayer distance) and yields approximately 4 K in $\kappa$-HgBr. 
	This value is considerably smaller than the values of the BCS case, 
typically 10$^{4}$-10$^{5}$ K \cite{Emery_1995}. 
	As the phase stiffness measures the macroscopic phase coherence in energy, 
this small value in $\kappa$-HgBr is consistent with the previously observed field resilience and 
strong fluctuations in superconductivity \cite{Suzuki_2022, Imajo_2021}. 
	From the penetration depth, the nominal Fermi temperature $T_{\textrm{F}}$ is 
evaluated through the relation 
$k_{\textrm{B}}T_{\textrm{F}}$ = 
$\pi\hbar^{2}d$/$\mu_{0}e^{2}\lambda_{\parallel}^{2}$ \cite{Uemura_1989} 
to be approximately 45 K, 
which gives a ratio, $T_{\textrm{c}}/T_{\textrm{F}}$, of $\sim$0.09. 
	This is a large value beyond the BCS framework, where $T_{\textrm{c}}$ is 
several orders of magnitude smaller than $T_{\textrm{F}}$. $\kappa$-HgBr is located in 
the unconventional pairing region in the $T_{\textrm{c}}$ versus 
$T_{\textrm{F}}$ plane (so-called Uemura plot), as shown in Fig. \ref{Fig4}(c) 
\cite{Uemura_1989, Uemura_1991, Melo_1993, Hazra_2019}.

	As seen above, the superfluid density takes a reduced value as if 
only the doped carriers condense into the superconducting state. 
	Such behavior is reminiscent of underdoped cuprates, 
in which the normal-state carrier density \cite{Ando_2004}, 
the Drude weight \cite{Basov_2005} and the superfluid density \cite{Uemura_1989, Panagopoulos_1999} 
are discussed to be governed by the doped carriers. 
	Theoretically, such a significant reduction in the superfluid density is suggested to occur 
when double occupancies of particles are strongly prohibited \cite{Lee_2006, WCLee_2008, Simard_2019}. 
	Previous transport studies on $\kappa$-HgBr at low pressures revealed that 
the mobile carriers are in a quantum critical state with a spin-liquid background 
\cite{Oike_2015, Suzuki_2022, Taniguchi_2007, Wakamatsu_2022}, 
which may bring about an incoherent part, preventing the entire spectral weight from 
contributing to the conduction and/or superconducting condensate. 
	It is argued for underdoped cuprates that some competing or hidden orders are responsible for 
the reduction in the superfluid density \cite{Chakravarty_2001} because various orders or 
anomalies appear in the spin and charge degrees of freedom competing with 
superconductivity \cite{Fradkin_2015, Keimer_2015}. 
	In this respect, $\kappa$-HgBr is distinct from cuprates in that 
no such competing orders have ever been observed, 
probably because a QSL free form order is the mother phase of the superconductivity. 
	Although in recent years the possibility of competing orders in doped triangular lattices 
has theoretically suggested in certain parameter spaces \cite{Zhu_2022, Wietek_2021}, 
the observation of a considerable reduction in the superfluid density without 
apparent competing orders in the present system would offer a focused target to 
diverse theoretical challenges with regard to the issue of missing superfluid spectral weight.
\\

%\section{ACKNOWLEDGMENT}
\begin{acknowledgments}
	This work was supported by Japan Society for the Promotion of Science (grant Nos. 18H05225, 19H01846, 20K20894, 20KK0060, 21K18144), JST SPRING (grant JPMJSP2108), and the Mitsubishi Foundation (grant 202110014). 
	Several experiments were performed using facilities of the Cryogenic Research Center, the University of Tokyo.
\end{acknowledgments}

%{Appendix}

$\\$

$^{\ast}$Corresponding author, e-mail: kanoda@ap.t.u-tokyo.ac.jp

\end{document}